\documentclass[twocolumn,aps,showpacs,prb]{revtex4}

\usepackage{amsmath,amssymb,graphics,epsfig,epstopdf,color,multirow,array,verbatim,ulem,braket,tabularx,graphicx}
\usepackage[colorlinks,linkcolor=blue,citecolor=blue,urlcolor=blue]{hyperref}
\usepackage{color}
\usepackage{graphicx}
\usepackage{epstopdf}
\usepackage{amsmath}
\usepackage{multirow}
\usepackage{ulem}
\usepackage{orcidlink}

\begin{document}

\title{Tetragonal-phase SnOFeSe: A possible parent compound of FeSe-based superconductor}

\author{Xiao-Xiao Man\orcidlink{0000-0002-5172-3554}$^1$}
\author{Pei-Han Sun\orcidlink{0000-0003-0177-8124}$^1$}
\author{Jian-Feng Zhang\orcidlink{0000-0001-7922-0839}$^2$}
\author{Zhong-Yi Lu\orcidlink{0000-0001-8866-3180}$^{1}$}
\email{zlu@ruc.edu.cn}
\author{Kai Liu\orcidlink{0000-0001-6216-333X}$^{1}$}
\email{kliu@ruc.edu.cn}

\affiliation{$^1$Department of Physics and Beijing Key Laboratory of Opto-electronic Functional Materials $\&$ Micro-nano Devices, Renmin University of China, Beijing 100872, China \\
$^2$Institute of Physics, Chinese Academy of Sciences, Beijing 100190, China}

\date{\today}

\begin{abstract}
Recent experiments have reported that inserting metal atoms or small molecules in between the FeSe layers of $\beta$-FeSe can significantly enhance the superconducting transition temperature. Here, based on first-principles electronic structure calculations, we propose a stable compound SnOFeSe by alternatively stacking the SnO and $\beta$-FeSe layers. The predicted SnOFeSe has the same tetragonal structure as the well-known FeAs-based compound LaOFeAs, meanwhile their electronic structures in the nonmagnetic state are quite similar. The magnetic ground state of SnOFeSe is predicted to be the dimer antiferromagnetic (AFM) state, which is energetically only 2.77 (2.15) meV/Fe lower than the trimer (dimer-trimer-dimer-trimer) AFM state, indicating that strong magnetic fluctuations might be induced via slight modulation. Interestingly, SnOFeSe is at the verge of metal-insulator transition in these low-energy magnetic states, hence bridging the metallic parent compounds of iron-based superconductors and the insulating ones of cuprate superconductors. With the reduced dimensionality, monolayer SnOFeSe also shows great similarities in the electronic and magnetic properties to its bulk phase. Given that SnOFeSe is adjacent to magnetic frustration and resembles LaOFeAs in both crystal and electronic structures, we suggest that SnOFeSe is a possible superconductor parent compound, which may provide a promising platform to study the interplay between magnetism and unconventional superconductivity in FeSe-derived materials.
\end{abstract}

%\pacs{}

\maketitle

\section{INTRODUCTION}

Iron-based superconductors have inspired intensive interests both experimentally and theoretically since their discovery in 2008~\cite{Iron SCs-1, Iron SCs-2, Iron SCs-3, Iron SCs-4, Iron SCs-5, Iron SCs-6, Iron SCs-7}. With the simplest crystal structure but the mysterious magnetic property among the iron-based superconductors ever found~\cite{Iron SCs-1, Iron SCs-2, Iron SCs-3, Iron SCs-4, Iron SCs-5, Iron SCs-6, Iron SCs-7}, the anti-PbO-type $\beta$-FeSe has become a prototypical system to study the physical properties of iron-based superconductors. The superconducting transition temperature \textit{T}$_\text{c}$ of bulk $\beta$-FeSe is about 8 K at ambient pressure~\cite{Iron SCs-2}, and numerous works, therefore, have been devoted to improve the \textit{T}$_\text{c}$ via high pressure~\cite{pressure-1, pressure-2}, chemical substitution~\cite{substitution}, gate voltage~\cite{gate}, or epitaxial film growth~\cite{film}. In addition, intercalation of metal atoms or molecules into $\beta$-FeSe is another effective approach to enhance the superconductivity. For instance, metal-intercalated FeSe compounds, \textit{A}$_\text{x}$Fe$_\text{2-y}$Se$_2$ (\textit{A} = Li, Na, K, Rb, Cs, Ca, Sr, Ba, Yb, Eu, Tl/K, or Tl/Rb), have shown the superconducting $\textit{T}_\text{c}$'s around 30 K~\cite{metal-1, metal-2, metal-3, metal-4, metal-5, metal-6, metal-7, metal-8, metal-9}. With the intercalation of Li$_x$(NH$_2$)$_y$(NH$_3$)$_{1-y}$~\cite{NH3}, (Li/Na)$_x$(NH$_3$)$_y$~\cite{Lei}, Li$_x$(C$_5$H$_5$N)$_y$~\cite{organic}, or Li$_{0.8}$Fe$_{0.2}$OH~\cite{LiFeOH-1, LiFeOH-3, LiFeOH-2} layers into $\beta$-FeSe, the \textit{T}$_\text{c}$'s can be significantly lifted above 40 K. The enhanced superconductivities in the FeSe-based compounds are believed to be intimately correlated with the modulation of electronic and magnetic properties of the FeSe layers.

\begin{figure}[!t]
\centering
\includegraphics[width=0.98\columnwidth]{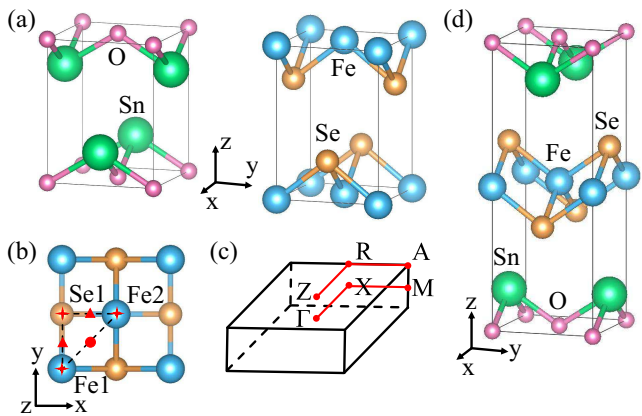}
\caption{(Color online) (a) Crystal structures of the tetragonal-phase SnO and $\beta$-FeSe. (b) Six typical stacking sites of the SnO layer on top of the FeSe layer, labeled by the red crosses, dot, and triangles. (c) Brillouin zone (BZ) and (d) crystal structure of bulk SnOFeSe. The high-symmetry $k$ points in the BZ are indicated by the red dots.}
\label{fig:1}
\end{figure}

We notice that the PbO-type SnO (space group: P4/\textit{nmm}) has the same tetragonal structure as the anti-PbO-type $\beta$-FeSe~\cite{SnO-1, SnO-2, SnO-3} [Fig. \ref{fig:1}(a)], although their metallic and nonmetallic elements are interchanged in atomic positions~\cite{SnO-1}. The experimental in-plane lattice constant of SnO ($a$ = 3.803 \text{\AA})~\cite{SnO-1} matches quite well with that of $\beta$-FeSe ($a$ = 3.765 \text{\AA})~\cite{Iron SCs-2}. It is thus very likely to form the 1111-type FeSe-based compound SnOFeSe by alternatively stacking the SnO and FeSe layers [Fig. \ref{fig:1}(d)], which is structurally identical to the FeAs-based compound LaOFeAs~\cite{Iron SCs-3}. Since SnO is reported to be a semiconductor with an indirect band gap of 0.7 eV~\cite{SnO gap-1, SnO gap-2}, the SnO layers can serve as insulating spacers in SnOFeSe and may make the FeSe layers more two-dimensional (2D) in electronic behavior. In addition, SnOFeSe has the equal valence electrons as LaOFeAs, thus it may resemble the latter in electronic properties and might also become superconducting via appropriate modulation~\cite{Iron SCs-3, Iron SCs-6, modulation-1, modulation-2, modulation-3, modulation-4, modulation-5}. Based on these preliminary inferences, the physical properties of the 1111-type FeSe-based compound SnOFeSe deserve further in-depth theoretical investigation.

By using first-principles electronic structure calculations, here we propose a stable intercalation compound SnOFeSe formed by inserting SnO layers into $\beta$-FeSe, which is isostructural to the well-known LaOFeAs. We find that the electronic structure of SnOFeSe in the nonmagnetic state is quite similar to that of LaOFeAs. And we determine the magnetic ground state of SnOFeSe to be a dimer antiferromagnetic state, in which the system is at the verge of metal-insulator transition. The monolayer SnOFeSe in the ultrathin limit has also been investigated in theory, which might be fabricated experimentally via epitaxial growth or mechanical exfoliation. Our calculation results suggest that SnOFeSe could be an ideal parent compound for studying the relationship between magnetism and unconventional superconductivity in FeSe-based superconductors.

\section{COMPUTATIONAL DETAILS}

To study the crystal structure, electronic structure, and magnetic properties of SnOFeSe, fully spin-polarized density functional theory (DFT) calculations were performed with the projector augmented wave (PAW) method~\cite{PAW-1, PAW-2} as implemented in the Vienna \textit{Ab initio} Simulation Package (VASP)~\cite{VASP-1, VASP-2, VASP-3}. The generalized gradient approximation (GGA) of the Perdew-Burke-Ernzerhof (PBE) type~\cite{GGA} was adopted for the exchange-correlation functional. The kinetic energy cutoff of the plane-wave basis was set to 520 eV. The DFT-D2 method~\cite{DFT D2-1, DFT D2-2} was used to account for the van der Waals (vdW) interaction in the layered materials~\cite{layered materials}. The $16 \times 16 \times 8$ and $4 \times 12 \times 8$ Monkhorst-Pack $\bf{k}$-point meshes were adopted to sample the Brillouin zones (BZs) of the unit cell and the supercell, respectively. The Fermi surface was broadened by the Gaussian smearing method with a width of 0.05 eV. The internal atomic positions and the cell parameters were fully optimized until the residual forces on all atoms were smaller than 0.01 eV/\text{\AA}. The energy convergence criterion was set to 10$^{-6}$ eV. For monolayer SnOFeSe, a vacuum layer larger than 20~\text{\AA} was utilized to eliminate the interaction between image slabs along the (001) direction. The interlayer binding energy \textit{E}$_\text{b}$ of SnOFeSe was calculated according to the formula \textit{E}$_\text{b}$ = \textit{E}$_\text{SnOFeSe}$ $-$ \textit{E}$_\text{FeSe}$ $-$ \textit{E}$_\text{SnO}$, where \textit{E}$_\text{SnOFeSe}$ is the energy of SnOFeSe, and \textit{E}$_\text{FeSe}$ (\textit{E}$_\text{SnO}$) is the energy of the FeSe (SnO) after removing the SnO (FeSe) layers from SnOFeSe. To check the dynamical stability, phonon spectra were calculated within the framework of density functional perturbation theory (DFPT) as implemented in the Quantum ESPRESSO (QE) package~\cite{QE}. For the thermal stability, \textit{ab initio} molecular dynamics simulations were carried out with the VASP package~\cite{VASP-1, VASP-2, VASP-3}. An \textit{NVT} ensemble with a temperature \textit{T} of 300 K controlled by the No$\acute{\text{s}}$e-Hoover thermostat was simulated. The total simulation time was set to 9 ps with a time step of 3 fs.

\section{RESULTS AND ANALYSIS}
\label{sec:Results}

\begin{figure}[!t]
\centering
\includegraphics[width=0.9\columnwidth]{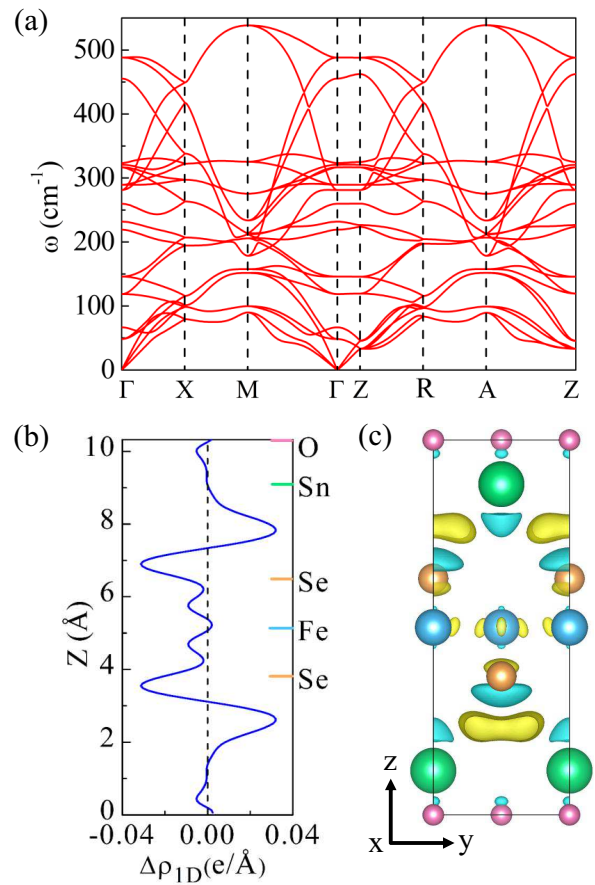}
\caption{(Color online) (a) Phonon dispersion of bulk SnOFeSe in the nonmagnetic (NM) state along the high-symmetry paths of the BZ [Fig.~\ref{fig:1}(c)]. (b) One-dimensional (1D) and (c) three-dimensional (3D) differential charge densities [$\Delta\rho$ = $\rho_\text{SnOFeSe} - \rho_\text{SnO} - \rho_\text{FeSe}$]. The atomic positions in panel (b) are marked by the color bars on the right axis. The yellow and cyan isosurfaces in panel (c) represent the electron accumulation and depletion areas, respectively. The isosurface value is set to $6.14 \times 10^{-4}$ $e/\text{\AA}^3$.}
\label{fig:2}
\end{figure}

The crystal structures of the tetragonal-phase SnO and $\beta$-FeSe (space group: P4/\textit{nmm}) are shown in Fig.~\ref{fig:1}(a). The experimental in-plane lattice constant of SnO is $a$ = 3.803 ~\text{\AA}~\cite{SnO-1}, which is only 1.0\% larger than that of $\beta$-FeSe ($a$ = 3.765 ~\text{\AA})~\cite{Iron SCs-2}. The rather small lattice mismatch between SnO and $\beta$-FeSe indicates that they might form the 1111-type FeSe-based compound SnOFeSe by alternatively intercalating the SnO and FeSe layers. In order to find out the most stable intercalation structure, we considered six nonequivalent stacking sites for the SnO layer on the FeSe plane [Fig.~\ref{fig:1}(b)], among which there are three top sites (Fe1 top, Fe2 top, and Se1 top) labeled by the red crosses, one hollow site labeled by the red dot, and two bridge sites (Fe1-Se1 bridge and Fe2-Se1 bridge) labeled by the red triangles. The concrete structures of SnOFeSe corresponding to the above stacking sites are displayed in Fig. S1 of the Supporting Information (SI)~\cite{SuppInfo}. After fully relaxing these structures in the nonmagnetic (NM) state, we found that the energetically most stable one [Fig.~\ref{fig:1}(d)] has the same structure as the 1111-type FeAs-based compound LaOFeAs~\cite{Iron SCs-3}. The calculated lattice constants of SnOFeSe are $a$ = $b$ = 3.736 ~\text{\AA} and $c$ = 10.321 ~\text{\AA}, in which the in-plane lattice constant is quite close to those of pristine SnO and $\beta$-FeSe~\cite{Iron SCs-2, SnO-1}, suggesting the small strain effect after the intercalation. The Brillouin zone (BZ) along with the high-symmetry $k$ points of SnOFeSe are schematically displayed in Fig.~\ref{fig:1}(c).

\begin{figure}[!t]
\centering
\includegraphics[width=0.97\columnwidth]{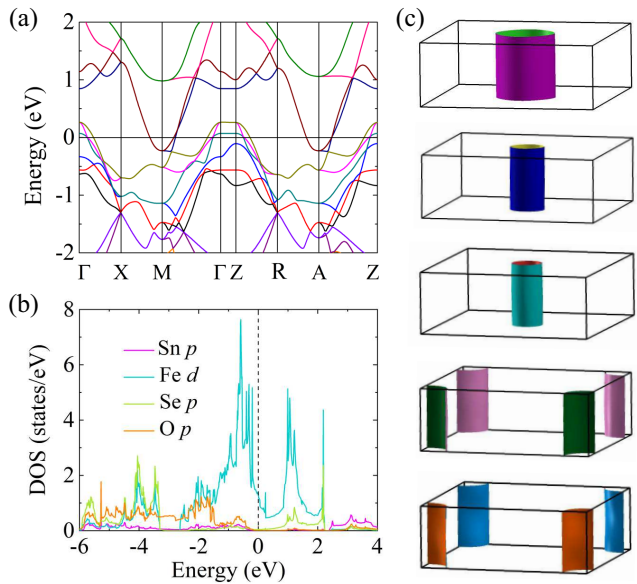}
\caption{(Color online) (a) Band structure along the high-symmetry BZ paths, (b) partial density of states (PDOS), and (c) five Fermi surface sheets for bulk SnOFeSe in the NM state. The Fermi energy is set to zero.}
\label{fig:3}
\end{figure}

We next examine the structural stabilities of SnOFeSe. The calculated phonon dispersion of SnOFeSe in the NM state along the high-symmetry paths of the BZ [Fig.~\ref{fig:1}(c)] is shown in Fig.~\ref{fig:2}(a), in which no imaginary phonon mode is observed, indicating the dynamical stability. Furthermore, we carried out \textit{ab initio} molecular dynamics simulations for the NM state of SnOFeSe, and obtained the time evolution of the free energy at 300 K as shown in Fig. S2(a) of the SI~\cite{SuppInfo}. From a snapshot at 9 ps [Figs. S2(b) and S2(c) in the SI~\cite{SuppInfo}], we can see that the structure remains intact without the bond breaking, suggesting that SnOFeSe is also thermodynamically stable at room temperature. The interlayer binding energy of SnOFeSe is calculated to be 112 meV/atom in the NM state, which is approximately two to three times that of graphite~\cite{graphite-1, graphite-2, graphite-3, graphite-4}. Such small interlayer binding energy indicates that SnOFeSe has a relatively weak interlayer interaction. To figure out the charge transfer between the SnO and FeSe layers in SnOFeSe, we then calculated the one-dimensional (1D) and three-dimensional (3D) differential charge densities as plotted in Figs.~\ref{fig:2}(b) and ~\ref{fig:2}(c), respectively. It can be seen clearly that there are some electrons accumulated in the interlayer region. These calculation results show that SnOFeSe is a dynamically and thermodynamically stable material that is very likely to be synthesized experimentally.

\begin{figure}[!t]
\centering
\includegraphics[width=0.97\columnwidth]{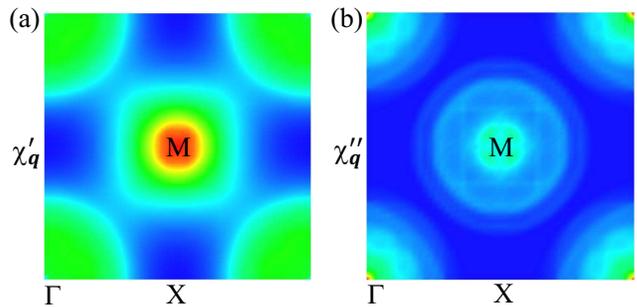}
\caption{(Color online) (a) Real and (b) imaginary parts of the electron susceptibility $\chi$ plotted in the $q$$_\text{z}$ = 0 plane for bulk SnOFeSe in the NM state.}
\label{fig:4}
\end{figure}

To investigate the electronic structure of SnOFeSe in the NM state, we calculated the band structure, the partial density of states (PDOS), as well as the Fermi surface (FS). From the band structure shown in Fig.~\ref{fig:3}(a), we can see that there are five bands crossing the Fermi level, indicating a metallic behavior of SnOFeSe. The overall band characteristics look very similar to those of LaOFeAs~\cite{LaOFeAs-Lu, LaOFeAs-Singh} [Fig. S3(a) in the SI~\cite{SuppInfo}], especially for the bands near the Fermi level. The calculated PDOS of SnOFeSe in Fig.~\ref{fig:3}(b) demonstrates that the Fe 3$d$ orbitals contribute most in the energy range from -2 to 2 eV around the Fermi level, which is also similar to that of LaOFeAs [Fig. S3(b) in the SI~\cite{SuppInfo}]. Figure~\ref{fig:3}(c) shows five FS sheets of SnOFeSe in the NM state, among which there are three hole-type pockets around the BZ center ($\Gamma$ point) and two electron-type pockets around the BZ corner (M point). The perfect cylindrical shape of these FS sheets indicates the prominent 2D feature of SnOFeSe. Based on the information of FS, we then calculated the electron susceptibility $\chi$($\bf{q}$). The real part $\chi^{\prime}$($\bf{q}$) shows a broad peak around the M point [Fig.~\ref{fig:4}(a)], suggesting the electronic instability of the NM state. Meanwhile, the imaginary part $\chi^{\prime\prime}$($\bf{q}$) [Fig.~\ref{fig:4}(b)] also has considerable intensity around the M point, which reflects the FS nesting that can be discerned intuitively by shifting the Fermi pockets from $\Gamma$ to M in Fig.~\ref{fig:3}(c). These results indicate that besides the crystal structure [Fig.~\ref{fig:1}(d)], the electronic properties of SnOFeSe in the NM state also resemble those of LaOFeAs~\cite{LaOFeAs-Mazin}.

\begin{table*}[!t]
\caption{\label{tab:I} Relative energies $\Delta$\textit{E} (in unit of meV/Fe) of the FM, checkboard AFM N$\acute{\text{e}}$el, stripe AFM, dimer AFM, trimer AFM, tetramer AFM, and dimer-trimer-dimer-trimer (di-tri-di-tri) AFM states with respect to the NM state for bulk SnOFeSe. The corresponding average local moments $\overline{M}$ (in unit of $\mu_{B}$) on Fe atoms are also listed.}
\begin{center}
\begin{tabular*}{16cm}{@{\extracolsep{\fill}} ccccccccc}
\hline \hline
 State&  NM&  FM&  N$\acute{\text{e}}$el&  stripe&  dimer&  trimer&  tetramer&  di-tri-di-tri\\
\hline
 $\Delta E$ &  0.00&  168.69&  -33.15&  -67.01&  -86.34&  -83.57&  -77.31&  -84.19\\
 $\overline{M}$ &   -&  2.34&  1.75&  1.94&  2.04&  2.02&  1.99&  2.01\\
 \hline \hline
\end{tabular*}
\end{center}
\end{table*}

\begin{figure}[!b]
\centering
\includegraphics[width=0.9\columnwidth]{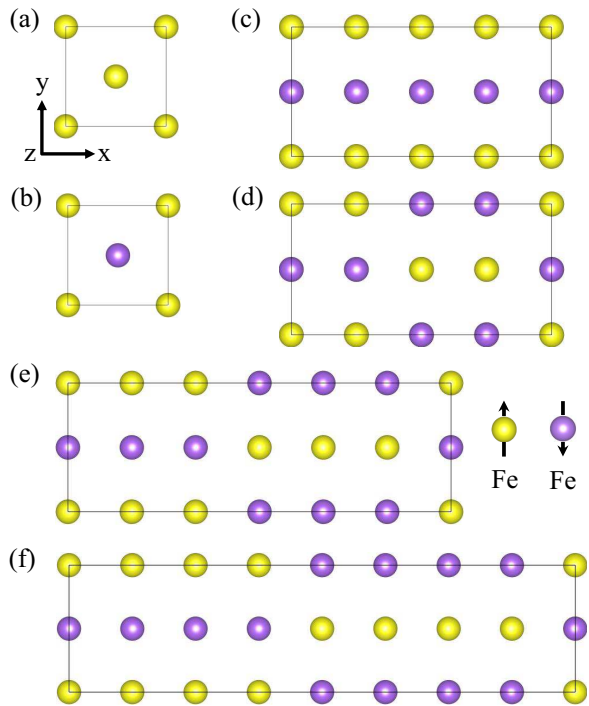}
\caption{(Color online) Sketches of six typical spin configurations for the Fe lattice in SnOFeSe: (a) ferromagnetic (FM) state, (b) checkerboard antiferromagnetic (AFM) N$\acute{\text{e}}$el state, (c) single-stripe (collinear) AFM state, (d) dimer AFM state, (e) trimer AFM state, and (f) tetramer AFM state. Here, the solid rectangles represent the supercells, while the purple and yellow balls represent the spin-up and spin-down Fe atoms, respectively.}
\label{fig:5}
\end{figure}

The existence of strong FS nesting in Fe-based superconductors may induce magnetic instabilities~\cite{nesting instability-1, nesting instability-2, nesting instability-3, nesting instability-4, nesting instability-5}, thus we have further studied the magnetic properties of SnOFeSe. In addition to the aforementioned NM state, we have investigated the ferromagnetic (FM) state and several typical AFM states including the checkerboard AFM N$\acute{\text{e}}$el, stripe AFM, dimer AFM, trimer AFM, and tetramer AFM states, whose spin configurations are schematically shown in Fig.~\ref{fig:5}. The calculated relative energies of these magnetic states with respect to that of the NM state are listed in Table~\ref{tab:I}. Clearly, the dimer AFM state possesses the lowest energy, being the magnetic ground state of SnOFeSe, which is similar to previous calculation results on $\beta$-FeSe~\cite{Gong FeSe, Liu FeSe}. It is worth noting that the energy of the trimer AFM state is quite close to and only 2.77 meV/Fe higher than that of the dimer AFM state. Inspired by our previous studies on $\beta$-FeSe~\cite{Liu FeSe}, we also investigated the AFM state combined by Fe spin dimers and trimers, such as the dimer-trimer-dimer-trimer (di-tri-di-tri) AFM state, which is energetically between the dimer and trimer AFM states and only 2.15 meV/Fe higher than the former one (Table~\ref{tab:I}). This turns out that with appropriate modulation there may exist magnetic fluctuations among these low-energy magnetic states (dimer, trimer, and their random combinations)~\cite{Liu FeSe}. Table~\ref{tab:I} also lists the average local moments on Fe atoms for these magnetic states from our calculation, whose values are comparable to the calculated local moments in $\beta$-FeSe~\cite{FeSe moment-1, FeSe moment-2}.

\begin{table*}[!t]
\caption{\label{tab:II} Relative energies $\Delta$\textit{E} (in unit of meV/Fe) of the magnetic states (AFM N$\acute{\text{e}}$el, stripe AFM, dimer AFM, trimer AFM, tetramer AFM, and di-tri-di-tri AFM) with respect to the NM state for monolayer SnOFeSe. The corresponding average local moments $\overline{M}$ (in units of $\mu_{B}$) on Fe atoms are also listed.}
\begin{center}
\begin{tabular*}{16cm}{@{\extracolsep{\fill}} cccccccc}
\hline \hline
 State&  NM&  N$\acute{\text{e}}$el&  stripe&  dimer&  trimer&  tetramer&  di-tri-di-tri\\
\hline
 $\Delta E$ &  0.00&  -41.12&  -73.39&  -90.04&  -88.23&  -83.42&  -89.30\\
 $\overline{M}$ &   -&  1.77&  2.01&  2.07&  2.08&  2.04&  2.05\\
 \hline \hline
\end{tabular*}
\end{center}
\end{table*}

\begin{figure}[!b]
\centering
\includegraphics[width=0.97\columnwidth]{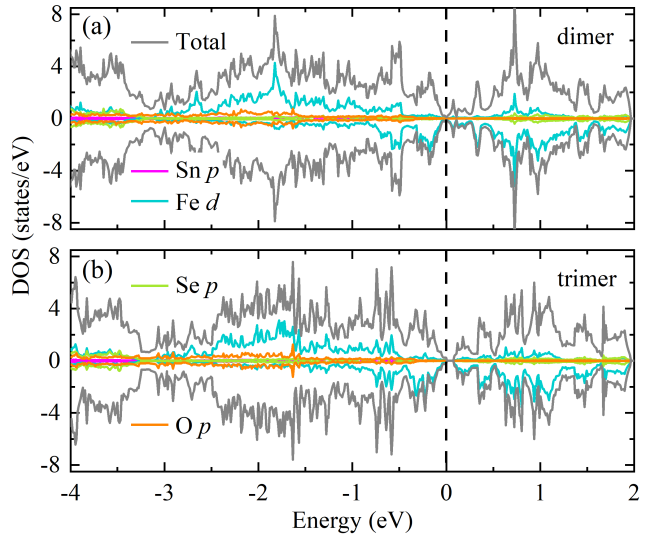}
\caption{(Color online) Total DOS and PDOS for (a) the dimer AFM and (b) the trimer AFM states of bulk SnOFeSe. The Fermi energy is set to zero.}
\label{fig:6}
\end{figure}

We next investigated the electronic structures of SnOFeSe in the above low-energy magnetic states. Figures  ~\ref{fig:6}(a) and  ~\ref{fig:6}(b) show the total and partial DOSs of SnOFeSe in the dimer and trimer AFM states, respectively. It can be seen that in these two AFM states, the Fe 3$d$ orbitals have the main contributions near the Fermi level. With a close inspection of the total DOS in the lowest-energy dimer AFM state, SnOFeSe shows a bad-metal character with very low density of states at the Fermi level. In comparison, there is a rather small energy gap ($\sim$0.07 eV) for SnOFeSe in the trimer AFM state. These results demonstrate that SnOFeSe is at the verge of metal-insulator transition when the magnetism is taken into account.

\begin{figure}[!b]
\centering
\includegraphics[width=0.9\columnwidth]{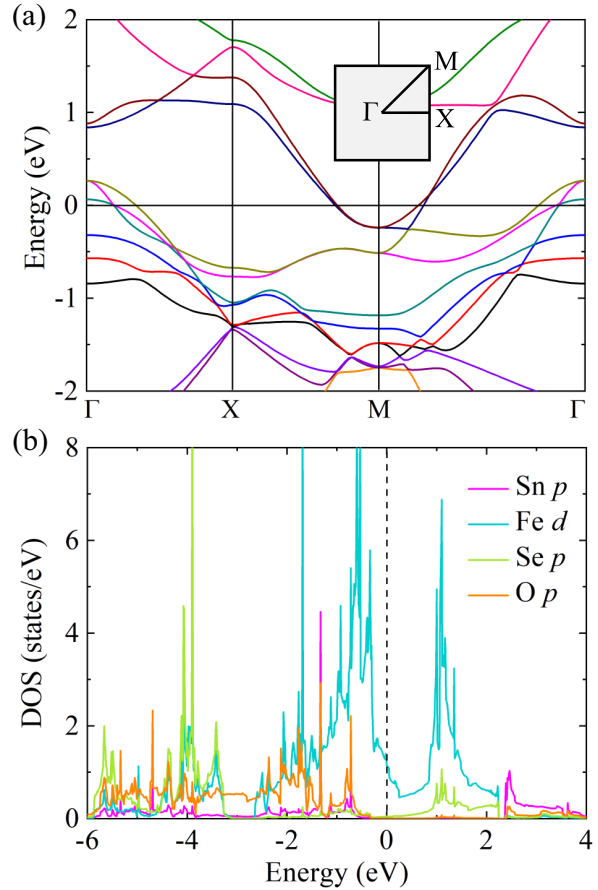}
\caption{(Color online) (a) Band structure along the high-symmetry paths of the BZ and (b) PDOS for monolayer SnOFeSe in the NM state. The Fermi energy is set to zero.}
\label{fig:7}
\end{figure}

In addition to the bulk phase, the monolayer form is also worth investigation for the layered materials. Likewise, we studied the dynamical stability as well as the electronic and magnetic properties of monolayer SnOFeSe. The calculated phonon dispersion of monolayer SnOFeSe in the NM state along the high-symmetry paths of the BZ is shown in Fig. S4 of the SI~\cite{SuppInfo}. A rather tiny imaginary frequency in the acoustic branch appears near the $\Gamma$ point, which is common for the calculated phonon spectra of 2D materials~\cite{PRB17, PRL09, PRB15}. It is not a sign of structure instability, but may originate from the difficulties in accurately calculating the rapid decaying interatomic forces~\cite{PRB09}. The band structure and PDOS of monolayer SnOFeSe in the NM state are shown in Figs.~\ref{fig:7}(a) and~\ref{fig:7}(b), respectively. The electronic band dispersion of monolayer SnOFeSe along the $\Gamma$-X-M-$\Gamma$ path [Fig.~\ref{fig:7}(a)] is similar to that of bulk phase [Fig.~\ref{fig:3}(a)], demonstrating the metallic behavior. Meanwhile, the DOS near the Fermi level is mainly contributed by Fe 3$d$ orbitals [Fig.~\ref{fig:7}(b)]. We have further studied several typical spin configurations (Fig.~\ref{fig:5}) for monolayer SnOFeSe. After full relaxation, the FM state converges to the NM state, suggesting that the former is unstable. The calculated relative energies of the AFM magnetic states with respect to the NM state as well as the local moments on Fe atoms for monolayer SnOFeSe are listed in Table~\ref{tab:II}. Like bulk SnOFeSe (Table~\ref{tab:I}), the dimer AFM state is also the lowest-energy spin configuration for the monolayer form. Besides, the energy of the trimer (dimer-trimer-dimer-trimer) AFM state is merely 1.81 (0.74) meV/Fe higher than that of the dimer AFM state, suggesting that via slight tuning there may also exist magnetic frustration among the low-energy magnetic states in monolayer SnOFeSe. From the calculated total DOSs of the low-energy magnetic states (Fig. S5 in the SI~\cite{SuppInfo}), we can see that in both the dimer and trimer AFM states, monolayer SnOFeSe possesses tiny band gaps. Overall, the electronic and magnetic properties of monolayer SnOFeSe show great similarities to those of the bulk form.

\section{Discussion and summary}
\label{sec:discussion}

The SnOFeSe compound formed by intercalating the SnO layers into the anti-PbO-type $\beta$-FeSe has the following advantages. First, the same tetragonal symmetry and the small lattice mismatch between SnO and $\beta$-FeSe allow for an ideal intercalation without structural reconstruction and hence the formation of a stable crystal. Second, according to our calculations in the NM state (Fig.~\ref{fig:3}), the 2D characteristics of the electronic structure of SnOFeSe are significantly enhanced compared with those of bulk $\beta$-FeSe. The improved two dimensionality due to the intercalation of insulating SnO layers facilitates the investigation of the intrinsic property of the FeSe layer. Third, SnOFeSe and $\beta$-FeSe show different magnetic behaviors and can thus be utilized for a comparative study. For $\beta$-FeSe, there is no long-range magnetic order at ambient pressure~\cite{no mag order} and the AFM order only emerges under certain high pressures~\cite{pressure-2, AFM order-1, AFM order-2}. Our previous calculations indicated that there exist quasidegenerate AFM states with tiny energy difference of 0.3 meV/Fe in $\beta$-FeSe, which are responsible for the absence of magnetic order~\cite{Liu FeSe}. Here, the magnetic ground state (dimer AFM state) of SnOFeSe is energetically 2.77 meV/Fe lower than that of the trimer AFM state, providing an opportunity to detect the AFM order of the FeSe layer at low temperature and ambient pressure. Hence, SnOFeSe would be an interesting platform to study both electronic structure and magnetism of the quasi-2D FeSe layers.

As to the superconducting properties, SnOFeSe has the same crystal structure as the famous parent compound of the 1111-type Fe-based superconductor, namely LaOFeAs. The electronic properties of SnOFeSe in the NM state, including band structure, density of states, and Fermi surface, are all similar to those of LaOFeAs. Moreover, both SnOFeSe and LaOFeAs have the antiferromagnetic ground states, which are often in close proximity to the superconducting phase~\cite{LaOFeAs-AFM}. Since previous experiments have shown that doping elements (such as F, Sr, Pb, Th, Ca/F, and Ce/F) in LaOFeAs can suppress the AFM order and achieve the superconducting \textit{T}$_\text{c}$'s up to $\sim$ 30 K~\cite{Iron SCs-3, Iron SCs-6, modulation-1, modulation-2, modulation-3, modulation-4, modulation-5}, we infer that the superconductivity in SnOFeSe can also be induced by appropriate modulations, such as element substitution~\cite{Iron SCs-4, Iron SCs-7, element-1}, external pressure~\cite{press-1, press-2}, ionic gating~\cite{gate, LiFeOH-2, ionic-1}, etc.

In summary, by using first-principles electronic structure calculations, we have predicted a stable intercalation compound SnOFeSe that is isostructural to LaOFeAs, and have systematically investigated its electronic and magnetic properties. We find that the electronic structures of SnOFeSe in the NM state show great similarities with those of LaOFeAs. In addition, the magnetic ground state of SnOFeSe is the dimer AFM state, which is energetically only 2.77 (2.15) meV/Fe lower than the trimer (dimer-trimer-dimer-trimer) AFM state, indicating that strong magnetic fluctuations may emerge with fine tuning. In the low-energy magnetic states, SnOFeSe is on the verge of metal-insulator transition, which bridges the respective metallic and insulating parent compounds of iron-based and cuprate superconductors. Further calculations reveal that the electronic and magnetic properties of monolayer SnOFeSe in the ultrathin limit resemble those of its bulk phase. In consideration of the similarities between SnOFeSe and LaOFeAs, we suggest that SnOFeSe is a promising parent compound to explore the FeSe-based superconductivity, which calls for future experimental verification.

\begin{acknowledgments}

We wish to thank J. G. Cheng, J. G. Guo, and X. L. Dong for helpful communications. This work was supported by the Beijing Natural Science Foundation (Grant No. Z200005), the National Natural Science Foundation of China (Grants No. 12174443 and No. 11934020), and the National Key R$\&$D Program of China (Grant No. 2019YFA0308603). Computational resources were provided by the Physical Laboratory of High Performance Computing at Renmin University of China.

\end{acknowledgments}

\end{document}